\documentclass[a4paper,fleqn,usenatbib]{mnras}
\pdfoutput=1
\usepackage{newtxtext,newtxmath}
\usepackage[T1]{fontenc}
\usepackage{ae,aecompl}

\usepackage{graphicx}	% Including figure files
\usepackage{amsmath}	% Advanced maths commands
\usepackage{amssymb}	% Extra maths symbols
\usepackage{xcolor}

\title{The velocity structure of Cygnus OB2}

\author[B. Arnold et al.]{
Becky Arnold,$^{1, 2}$\thanks{rebeccajasmi@umass.edu}
Simon P. Goodwin$^{1}$ and
Nick J. Wright$^{3}$
\\
$^{1}$Department of Physics and Astronomy, University of Sheffield, Sheffield S3 7RH, UK\\
$^{2}$Department of Astronomy, University of Massachusetts, Massachusetts, 01003, USA\\
$^{3}$Astrophysics Group, Keele University, Keele, Staffordshire, ST5 5BG, UK
}

\date{Accepted XXX. Received YYY; in original form ZZZ}

\pubyear{2019}

\begin{document}
\label{firstpage}
\pagerange{\pageref{firstpage}--\pageref{lastpage}}
\maketitle

\begin{abstract}
  The kinematic structure of the Cygnus OB2 association is investigated.
  No evidence of
  expansion or contraction is found at any scale within the
  region. Stars that are within $\sim$ 0.5 parsecs of one another are found
  to have more similar velocities than would be expected by random chance,
  and so it is concluded that velocity substructure exists on these scales.
  At larger scales velocity substructure is not found. We suggest that
  bound substructures exist on scales of $\sim$ 0.5 parsecs, despite the
  region as a whole being unbound. We further suggest that any
  velocity substructure that existed on scales $>$ 0.5 parsecs has been erased. The results of this study are then compared to
  those of other kinematic studies of Cygnus OB2.
\end{abstract}


%%%%%%%%%%%%%%%%%%%%%%%%%%%%%%%%%%%%%%%%%%%%%%%%%%

%%%%%%%%%%%%%%%%% BODY OF PAPER %%%%%%%%%%%%%%%%%%

\section{Introduction} \label{introduction}

Star forming regions are the focus of a great deal of scientific
interest, and for good reason. They inform our
understanding of how stars are born and how their environments evolve.
Their study is also vital for our comprehension of the conditions that planets form in,
and the type/number of planets which may exist in the universe.

Spatial and dynamical structure are perhaps the
most important aspects defining a star forming region,
but can be difficult to interpret. A number
of statistical methods have been developed to quantify different aspects
of the spatial structure of these regions \citep{Alison09a, Cartwright09, Maschberger11,Buckner19}, but they don't touch upon its velocity structure.
However the Velocity Structure Analysis Tool (VSAT) \citep{Arnold19} does, and it is used in this
paper to investigate the velocity structure of Cygnus OB2, which has previously
had relatively little statistical kinematic analysis.

Cygnus OB2 lies at a distance of approximately 1400 parsecs
\citep{Hanson03, Rygl12, Berlanas19}, and has an estimated stellar mass of
order 10$^4$ $M_\odot$ \citep{Drew08, Wright10}.
Estimates of the region's age vary, for example \citet{Massey95} find an age of
1-3 Myr and \citet{Wright15a} find an age up up to 7 Myr with star formation
peaking 4-5 Myr ago. Given a number of estimates it seems relatively certain
that the age of the region lies somewhere between 1 and 7 Myr.

This region is chosen because it has been extensively studied,
\citep{Massey91, Knodlseder00, Comeron08, Kiminki15, Roquette17, Berlanas18, Berlanas19}
meaning there is a large amount of observational data already
available. There have also been studies focused on its spatial and kinematic structure
\citep{Wright14, Wright16, Winter19}. This is useful as it
allows findings relating to the kinematic structure of the region which are
achieved using different techniques to be compared to see if they are consistent.

The structure of this paper is as follows. In section \ref{methods}
the methods used to collect and analyse the data are outlined. In
section \ref{results} the results of the analysis are reported and
in section \ref{discussion} they are discussed.
Finally in section \ref{conclusions} the conclusions drawn from the
results are summarised.

\section{Methods} \label{methods}

\subsection{Data collection} \label{data_collection}

For this work we use the X-ray selected sample of Cygnus OB2 members presented
by \citet{Wright09} for the central portion of the association. X-rays provide
a largely unbiased diagnostic of youth that is effective for separating young
association members from older field stars, and \citet{Wright10} made further
efforts to identify and remove foreground contaminants from the sample. Proper
motions for these stars were derived by \citet{Wright16} as part of the DANCe
(Dynamical Analysis of Nearby Clusters) project \citep{Bouy13}.

The sample
includes many of the known high-mass O-type stars in Cygnus OB2 with masses
up to 100 $M_\odot$, as well as low-mass stars down to ~0.1 $M_\odot$. The sample
is estimated to be mostly complete to $\sim$ 0.8 $M_\odot$ \citep{Wright14}. In contrast
only around 70 per cent of these sources appear in the $Gaia$ survey results and
approximately half of those do not survive $Gaia$'s recommended astrometric quality cuts \citep{Arenou18, Lindegren18a, Lindegren18b}.
This is because the sources are mostly fainter than is optimal for detection by $Gaia$.

\begin{figure*}
  \includegraphics[width=\textwidth,angle=270,origin=c]{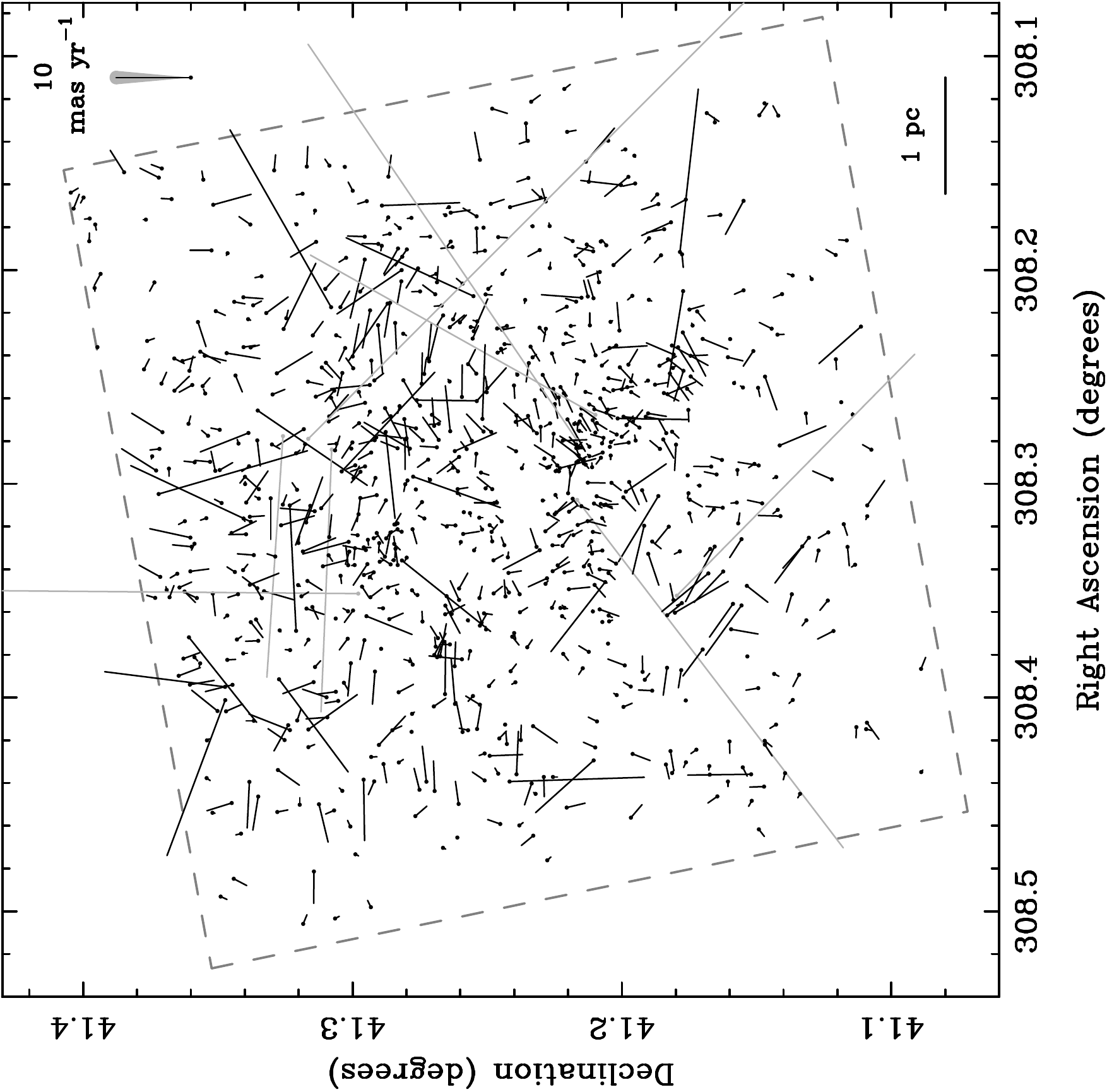}
  \caption{A plot of the data collected from the Cygnus OB2 region. The survey area of the X-ray observations of \citet{Wright09} is outlined by a grey dashed line. Each star is represented by a dot with a line. The position of the dot indicates the position of the star in RA  ($x$-axis) and Dec ($y$-axis). The length and direction of the line coming from each dot indicates that star's velocity vector. Stars which are removed from the sample (as discussed in section \ref{data_collection}) are shown by grey dots and vectors. In the top right hand corner is a 10 mas yr$^{-1}$ (equal to 66 km s$^{-1}$ at 1.4 kpc) vector with a grey cone outlining the median velocity uncertainty for the dataset.}
  \label{plot_of_region}
\end{figure*}

The dataset used in this paper can be
seen in Fig. \ref{plot_of_region}. In this figure each star is shown as a dot with
a line. The location of the dot indicates the location of the star and the length
and direction of the line indicates the star's velocity. In the top right of the
figure a velocity vector of 10 mas yr$^{-1}$ is shown with the median velocity uncertainty
of the dataset shown by a grey cone. This is done to give a visual representation of
the uncertainties on the velocity data.

Visual inspection of the
figure shows a small number of stars with velocities far greater than most. To
investigate this we conduct an exploratory analysis of the data. For additional details relating
to the origins of this dataset see \citet{Wright16}.

\subsection{Exploratory analysis} \label{exploratory analysis}

\begin{figure}
  \includegraphics[width=\columnwidth]{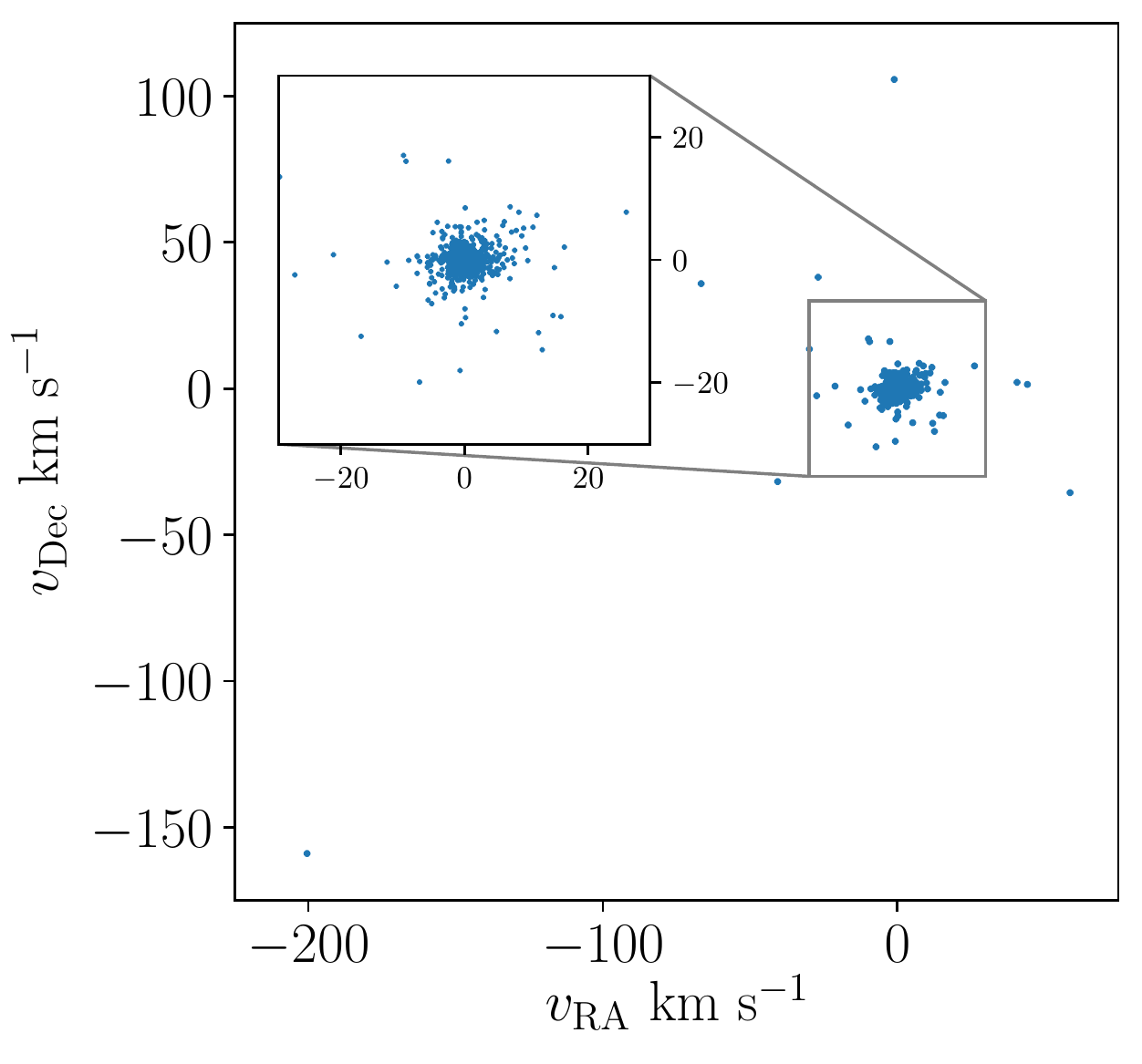}
  \caption{The RA and Dec stellar velocities plotted against one another.}
  \label{v_RA_vs_v_dec}
\end{figure}

\begin{figure}
  \includegraphics[width=\columnwidth]{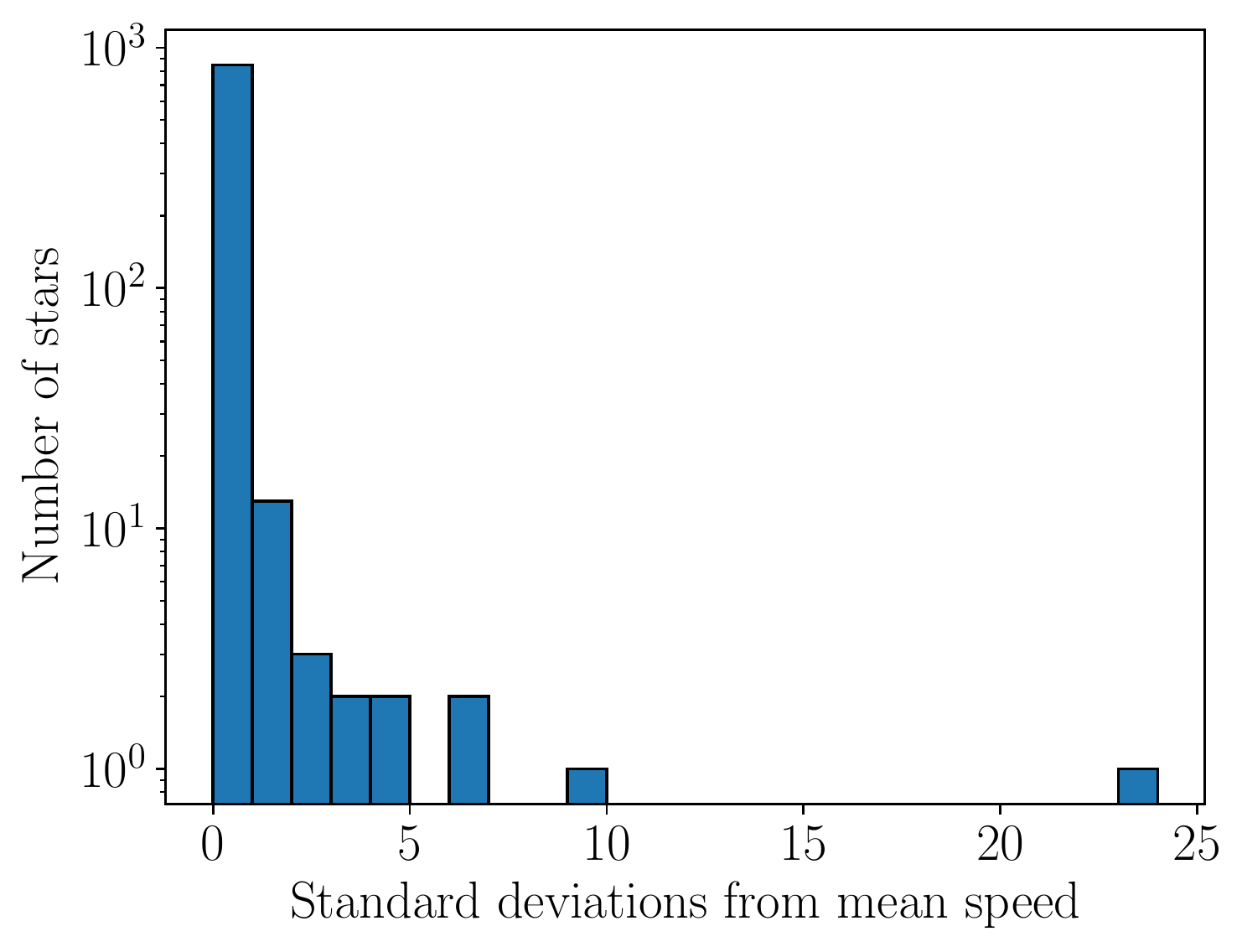}
  \caption{A histogram of the number of standard deviations stellar speeds are from the mean stellar speed in the Cygnus OB2 dataset. On the $x$-axis is number of standard deviations away from the mean. On the $y$-axis is the number of stars in each bin of the histogram. In order to ensure bins with a small number entries remain clearly visible the $y$-axis is logarithmic.}
  \label{sigma_clipping}
\end{figure}

Fig. \ref{v_RA_vs_v_dec} shows a scatter plot of the stellar velocities in the reference frame of the region. As was
the case for Fig. \ref{plot_of_region} it is visually apparent there are a small
number of outliers. Additionally Fig. \ref{sigma_clipping} shows a histogram
with the stars binned by the number of standard deviations their speed is
from the mean stellar speed of the dataset, and from this it is clear there is a small number
of stars which have speeds many standard deviations in excess of the mean.
Note that in Fig. \ref{sigma_clipping} the $y$-axis is
logarithmic in order to make bins with few entries more easily visible.

The presence of extreme velocity outliers is also noted in \citet{Wright16}
where this dataset was first presented. That work concludes that these outliers
are most likely due to:
\begin{itemize}
\item background/foreground stars that have been mistaken for members of Cygnus OB2;
\item stars ejected from Cygnus OB2 by dynamical events such as the disruption of binary systems;
\item stars now in the region of Cygnus OB2 but originating in nearby
  star clusters/associations which dispersed.
\end{itemize}

Regardless of their origin it is clear
there is a small number of the 873 stars with very different velocities to
the rest of the sample. Further, it seems unlikely that they are representative of
the region's underlying velocity structure. Due to this, stars with speeds
more than three standard deviations from the mean stellar speed in the
dataset (a total of eight stars) are removed. This removal has very little
impact on the results, and the effects it does have are discussed later.
The removed stars are shown in grey in Fig. \ref{plot_of_region} and outside the grey inset in
Fig. \ref{v_RA_vs_v_dec}

\begin{figure}
  \includegraphics[width=\columnwidth]{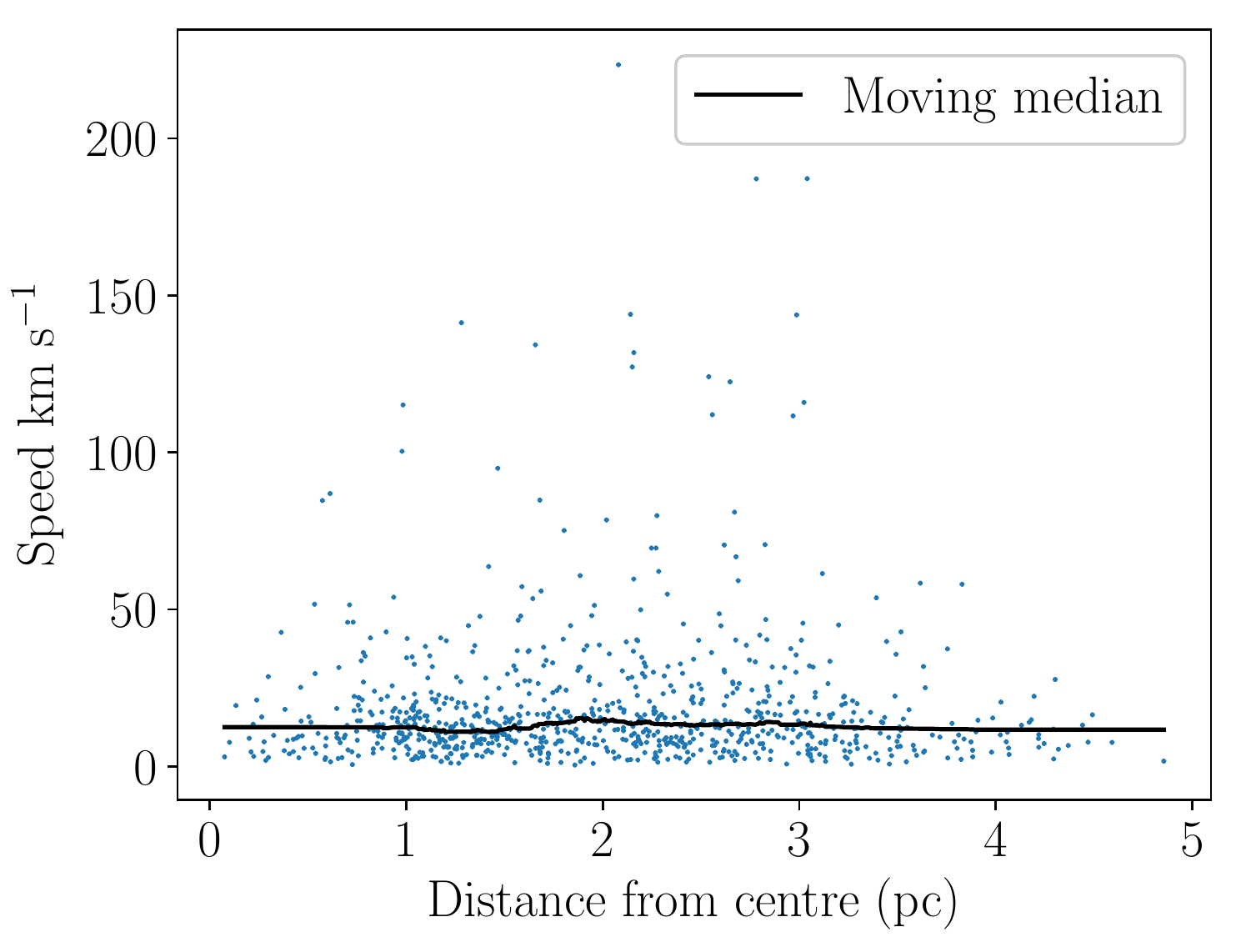}
  \caption{A scatterplot of stellar speeds plotted against their distance from
  the centre of the region. The moving median for this data is shown by
  a black line.}
  \label{dist_vs_speed}
\end{figure}

In Fig. \ref{v_RA_vs_v_dec} velocities appear to be centrally concentrated but,
particularly in the inset, there is no further obvious velocity structure apparent.
To further investigate the velocity structure we plot stellar speeds against their
distances from the centre, see Fig. \ref{dist_vs_speed}. The centre is defined
here as the mean position of the stars.

There appears to be a peak in this distribution of points towards the middle of
the plot but it is difficult to assess meaningfully by eye. To better understand it
the moving median is calculated and is shown by a black line on Fig.
\ref{dist_vs_speed}. Upon close examination the moving median appears slightly elevated at distances between $\sim$ 2 and 3 pc, indicating stars with high speeds are preferentially
located at moderate distances from the centre of the field of view. That said, this effect is extremely slight and the function is largely flat
indicating that any relationship between speed and position is weak.

\subsection{Data analysis} \label{data_analysis}

The VSAT method \citep{Arnold19} is applied to analyse the
velocity structure of the dataset in more detail. In brief the method is as follows:
for every possible pair of stars
this method calculates the distance between them ($\Delta r$)
and their velocity difference ($\Delta v$). To clarify, here a `pair of stars'
does not necessarily refer to a binary system, but just to any two stars in
the region. The pairs are then binned by
$\Delta r$ and within each bin the average $\Delta v$ is calculated.
Finally these are plotted against each other.
The uncertainties on the velocities are propagated when calculating each $\Delta v$, and these errors
are used to weight the average. The impact of stochasticity on these uncertainties is also incorporated, see
\citet{Arnold19} for full details of how these calculations are performed.

The VSAT method is applied twice, each time using a different definition of the
velocity difference between two stars $\Delta v$. The first case is referred to as the
magnitude definition, $\Delta v_{\rm M}$, and it is
defined as the magnitude
of the difference between the two star's velocity vectors. Therefore for stars $a$
and $b$ $\Delta v_{\rm M}$ is calculated as:

\begin{equation} \label{dv_orig}
  \Delta v_{a\!b \rm M} ~=~ \lvert \mathbf{v}_{a} - \mathbf{v}_{b} \rvert
\end{equation}

This definition is particularly useful as a raw measure
of how similar/different stellar velocity vectors are.

The second way $\Delta v$ is defined is as the time differential of
$\Delta r$, i.e. the rate at which the distance between the stars is changing.
This is referred to as the directional definition, $\Delta v_{\rm D}$.
For stars $a$ and $b$ this is

\begin{equation}
  \Delta v_{a\!b \rm D} ~=~ \frac{(\mathbf{r}_{a} - \mathbf{r}_{b})\boldsymbol{\cdot}(\mathbf{v}_{a} - \mathbf{v}_{b})}{\Delta r_{a\!b}}.
\end{equation}

In this definition if the stars are moving away from each other $\Delta r$
increases so $\Delta v_{\rm D}$ is positive, and if the stars are moving towards each
other $\Delta r$ decreases and $\Delta v_{\rm D}$ is negative. This definition is particularly useful
for studying if regions or sub-regions are undergoing expansion/collapse. It
as also useful for studying the relative motions of different substructures within
a region. Both of these things are helpful in discerning the dynamical state of
a region and its history.

In order to facilitate the calculation of these parameters the data is converted from RA, Dec,
and proper motions into a Cartesian coordinate scheme. The \textsc{\small{}skycoord} class in the Python package \textsc{\small{}astropy}
is used to perform this conversion.

\section{Results} \label{results}

\subsection{Magnitude definition: $\Delta v_{\rm M}$} \label{results_mag}

\begin{figure}
  \includegraphics[width=\columnwidth]{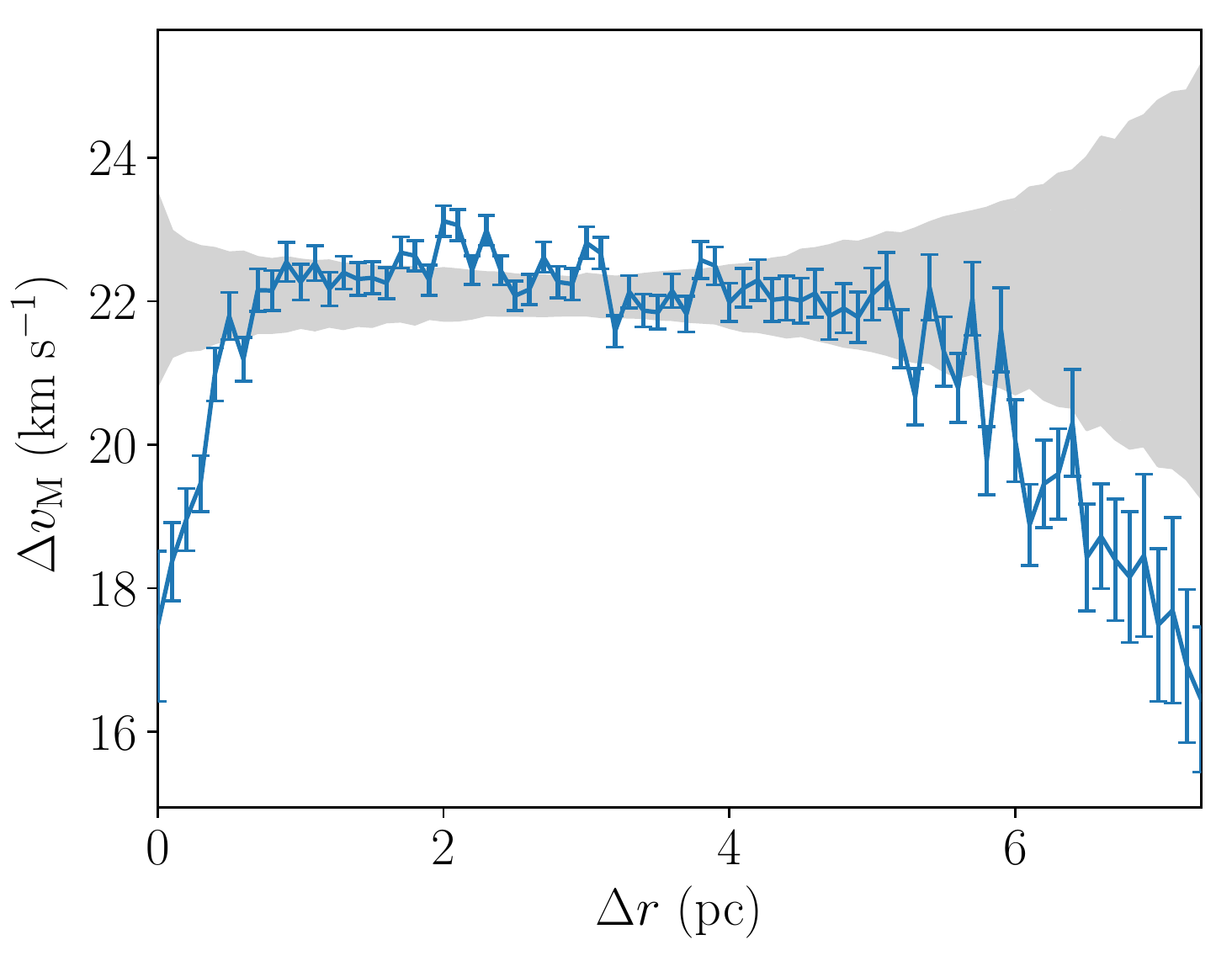}
  \caption{The velocity structure of Cygnus OB2 as determined by the VSAT method \citep{Arnold19} (blue line). The area containing the central 1$\sigma$ of the results of 1000 randomised cases is shown in grey. The $x$-axis shows the separation between stars in parsecs $\Delta r$, and on the $y$-axis is the average velocity difference of stellar pairs as defined by the magnitude definition of velocity difference, $\Delta v_{\rm M}$, in km s$^{-1}$.}
  \label{dv_mag}
\end{figure}

The VSAT method is applied to the dataset using the
magnitude definition $\Delta v_{\rm M}$. The results are shown by the blue
line in Fig. \ref{dv_mag}. On the $x$-axis of this figure is the distance between
stars in parsecs, $\Delta r$, and on the $y$-axis is the average magnitude
of the difference between stellar velocity vectors, $\Delta v_{\rm M}$, in km s$^{-1}$.

To demonstrate the degree of statistical noise the velocity vectors are
randomly swapped between stars to remove any velocity structure.
The method is re-run and the results recorded. This is done 1000 times.
The area containing the central
1$\sigma$ of results is shown by a shaded grey region in Fig. \ref{dv_mag}.
Features in the non-randomised results (blue line) on scales smaller
than the spread of these randomised results (grey area) are not significant
as they are within the fluctuations in measured velocity structure due to
pure stochasticity.

Inspection of Fig. \ref{dv_mag} shows a significant dip at scales $<$ 0.5 parsecs\footnote{This feature is not visible if stars that are sigma clipped in section \ref{data_collection} are included. This is because, due to their locations, these stars feature in a disproportionately large number of low $\Delta r$ pairs, and the $<$ 0.5 parsec scale is particularly badly impacted. The sigma clipped stars have very different velocities to the rest of the dataset so their over representation causes an artificial increase in $\Delta v_{\rm M}$. This explains their impact on the $<$ 0.5 parsec feature.}.
This means that stars closer to each other than this tend to
have similar velocity vectors relative to the velocities across the whole region studied.

Between $\sim$0.5 - 5.5 parsecs the measured velocity structure (blue) is almost
completely flat; there is no change in the magnitude of the difference
between stellar velocity vectors as a function of how far apart they
are. Further, the small fluctuations from flatness that are observed
are almost entirely within the 1$\sigma$ bounds of those resulting from the regions where
velocity structure has been removed by random shuffling
(grey area). Therefore we conclude that there does not appear to be any
velocity structure at these scales.

For $\Delta r >$ 5.5 parsecs
$\Delta v_{\rm M}$ dips again. However, the results
at these large scales are noisy and have high uncertainties. This is
largely due to the fact that only stars on the outermost edges of the dataset
are far enough apart to have such high $\Delta r$. Because of the morphology of the region which has
lower stellar density towards its outskirts this means that
there are fewer stellar pairs to populate these bins at large separations, and the results
at $\Delta r \gtrsim 5.5$ parsecs are highly dependent on the
exact velocity vectors of a small number of stars. Therefore although
the measured `structure' exceeds the bounds of the results with randomised
velocity structures we cannot confidently determine whether
velocity structure is or is not present at $\Delta r$ beyond 5.5 parsecs.

\subsection{Directional definition: $\Delta v_{\rm D}$} \label{results_dir}

\begin{figure}
  \includegraphics[width=\columnwidth]{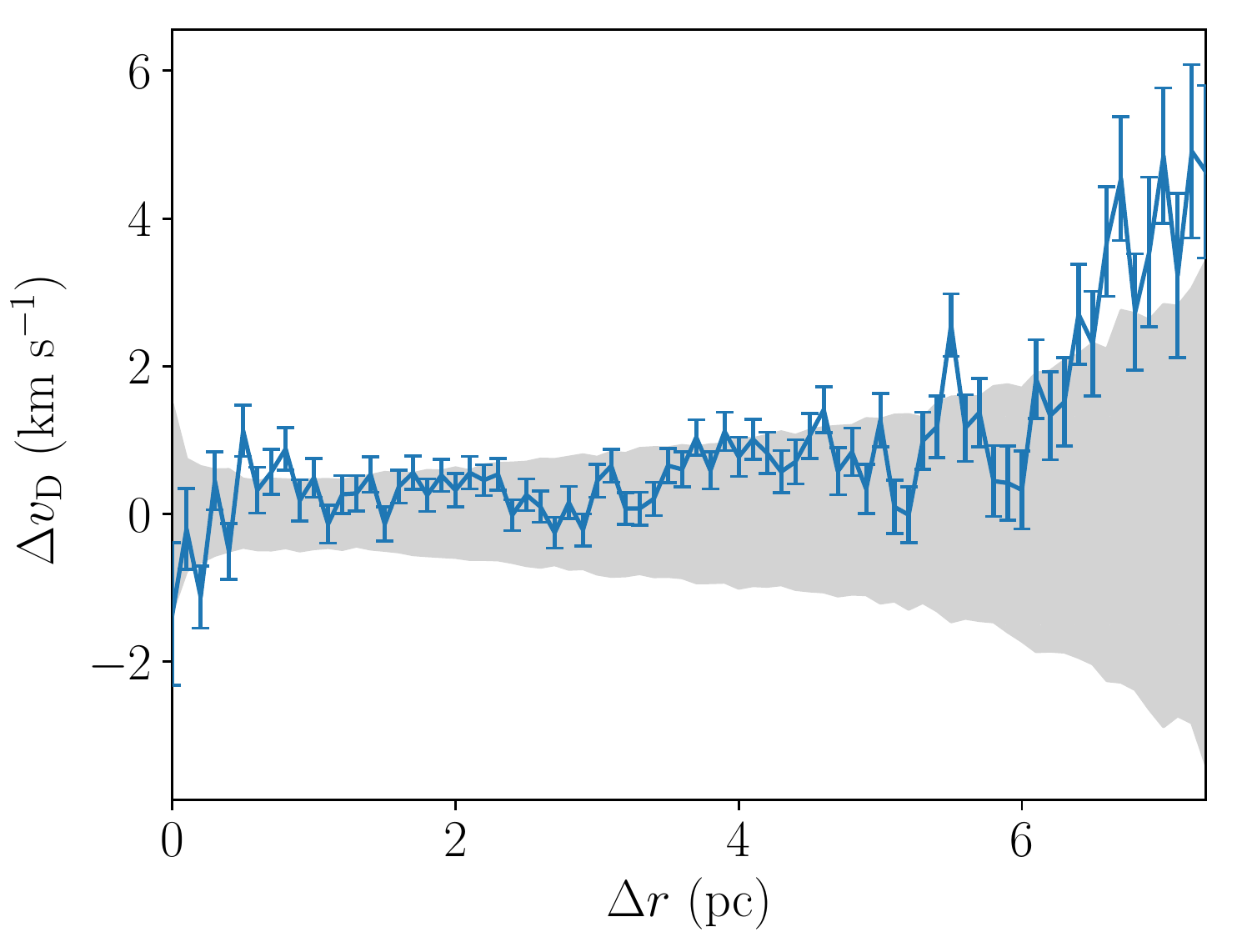}
  \caption{The velocity structure of Cygnus OB2 as determined by the VSAT method \citep{Arnold19} (blue line). The area containing the central 1$\sigma$ of the results of 1000 randomised cases is shown in grey. The $x$-axis shows the separation between stars in parsecs $\Delta r$, and on the $y$-axis is the average velocity difference of stellar pairs as defined by the directional definition of velocity difference, $\Delta v_{\rm D}$, in km s$^{-1}$.}
  \label{dv_dir}
\end{figure}

The VSAT method is now applied using the directional definition of
the velocity difference, $\Delta v_{\rm D}$. Recall that in this definition
the more rapidly stars tend to move away from each other the more positive
$\Delta v_{\rm D}$ is, and the more rapidly stars tend to move towards
each other the more negative it is. The results are shown by
the blue line in Fig. \ref{dv_dir}\footnote{These results are virtually
  indistinguishable from the results if the eight stars removed in section
  \ref{data_collection} are included. The only impact is minor changes to
  the exact fluctuations of the lines in Fig. \ref{dv_dir}, but the same
  overall trend is observed. This is as expected: if the clipped stars are truly not members of Cygnus OB2 they should have more or less random directions, so no net impact on the results of the directional definition $\Delta v_{\rm D}$.}. As in Fig. \ref{dv_mag}
the $x$-axis shows distance between pairs of stars, $\Delta r$,
in parsecs. The directional velocity difference $\Delta v_{\rm D}$
is given on the $y$-axis in km s$^{-1}$.
As in section \ref{results_mag}
velocity vectors are then randomly shuffled between stars to remove any
velocity structure, and the VSAT method is re-applied. This is done 1000 times.
The 1$\sigma$ boundary of the results is
plotted in grey on Fig. \ref{dv_dir}. This is done to give an
idea of the amplitude of apparent velocity structures which in fact result
from statistical noise.

The velocity structure of the region as measured using
the directional definition of velocity difference does not
fluctuate a great deal as a function of $\Delta r$.
The blue line in Fig. \ref{dv_dir} is largely flat, and
the results are consistent with $\Delta v_{\rm D} = 0$ (no net expansion or contraction)
for $\Delta r < 3.5$ parsecs. For $\Delta r$ between 3.5 and 6 parsecs $\Delta v_{\rm D}$
is only very slightly above 0, and stays almost entirely within the bounds of the
randomised cases with no velocity structure.

At $\Delta r > 6$ parsecs the results show a positive correlation
between $\Delta r$ and $\Delta v_{\rm D}$. Only stars on opposite sides of the dataset
and close to its edges are far enough apart to populate these high $\Delta r$ bins.
Additionally, as stated, positive $\Delta v_{\rm D}$ indicates a stellar pair is moving
apart. Therefore this result implies that stars towards the fringes of the dataset
are moving outwards, i.e that the region is expanding from it edges. However, for reasons
discussed in section \ref{results_mag} VSAT is not reliable beyond
$\Delta r =$ 5.5 parsecs for this dataset because of low number statistics.
Therefore an additional test is conducted to verify whether this apparent expansion is real.

First all the stars that are present in $\Delta r > 6$ parsec bins (we will call
these fringe stars) are identified, and how many times they appear in those
bins is recorded. Next the centre of the dataset is determined by taking the
average position of all stars. The direction of each fringe star's velocity relative
to the centre of the dataset is then calculated, with 0$^\circ$ indicating
the star is moving radially outwards, 90$^\circ$ that it is moving tangentially,
and 180$^\circ$ that is moving radially inwards. If fringe stars are moving systematically
outwards then we expect their mean angle to be $<$ 90$\pm 2.90^\circ$. Note 2.90$^\circ$ is
the expected standard deviation of the mean of a uniform distribution between 0 and 180$^\circ$
given the number of datapoints.

The mean angle of the fringe star's velocity directions is calculated. This mean is
weighted by how many times each star appears in $\Delta r > 6$ parsec bins, as recorded
earlier. The mean is 88.00$^\circ$, however this is within the expected standard
deviation of 2.90$^\circ$ so therefore is not significant. From this we conclude
there is no evidence of expansion or contraction at any scale in Cygnus OB2.

We emphasise that these findings do not constitute a null result. The fact that the data does not
convincingly deviate from the range occupied by the randomised cases in Fig. \ref{dv_dir}
demonstrates that if there is any velocity structure in the dataset according to the $\Delta v_{\rm D}$
definition then it is so small as to be indistinguishable from random noise.
As such the principle finding of this section is that if there is any velocity structure
present in Cygnus OB2 is is less significant than random fluctuations.

This raises the question of how significant would the region's expansion have to be in
order to be distinguishable from random fluctuations. In order to test this regions with increasingly
significant expansion are simulated, and VSAT is applied. The regions are simulated by first
drawing random positions from a uniform distribution and velocities from a Gaussian distribution with a width of 1, both centred on zero.
Note the artificial regions are generated with the same number of stars as the Cygnus OB2 dataset.
Random radial velocity components are then added to each star. The magnitudes of these velocities are also
drawn from a Gaussian with a width of 1 and a mean of $R_{\rm {sig}}$, which we define as the significance
of the outwards expansion. This can be thought of as the ratio of systematic expansion to randomness. Note that
if the number drawn for the radial magnitude is negative the radial component added will point inwards rather than outwards.

$R_{\rm {sig}}$ is increased in increments of 0.01. For each $R_{\rm {sig}}$ 100 regions are generated,
VSAT is applied, and their $\Delta v_{\rm D}(\Delta r)$ is recorded. Results at $\Delta r >$ 1 are
excluded as they rely on the `corners' of the uniform distribution, and so rely on a smaller number of datapoints
than results at smaller $\Delta r$ so are less reliable. This is analogous to the discounting of results
at $\Delta r >$ 5.5 parsecs in the Cygnus OB2 dataset's results.

The $\Delta v_{\rm D}(\Delta r)$s of the expanding regions are then compared to simulated regions
that are not expanding in order to determine if VSAT can differentiate them; another 100 regions
that do not have systematic expansion are generated and VSAT is applied. For each $\Delta r$
the range of $\Delta v_{\rm D}$s that contain the central
1$\sigma$ of results is recorded. These limits encapsulate the expected 1$\sigma$
fluctuation in $\Delta v_{\rm D}(\Delta r)$ due to stochasticity
(analogous to the grey regions in Fig. \ref{dv_mag} and Fig. \ref{dv_dir}).
The mean fraction of $\Delta v_{\rm D}(\Delta r)$ datapoints outside of these limits for the expanding regions is recorded.
As expected this fraction increases with $R_{\rm {sig}}$, and increases beyond 1$\sigma$ significance
when $R_{\rm {sig}}$ is 0.07. Therefore VSAT can identify expanding regions if the ratio of systematic expansion to
random velocity dispersion is $\gtrsim$ 0.07.

\begin{figure*}
  \includegraphics[width=\textwidth]{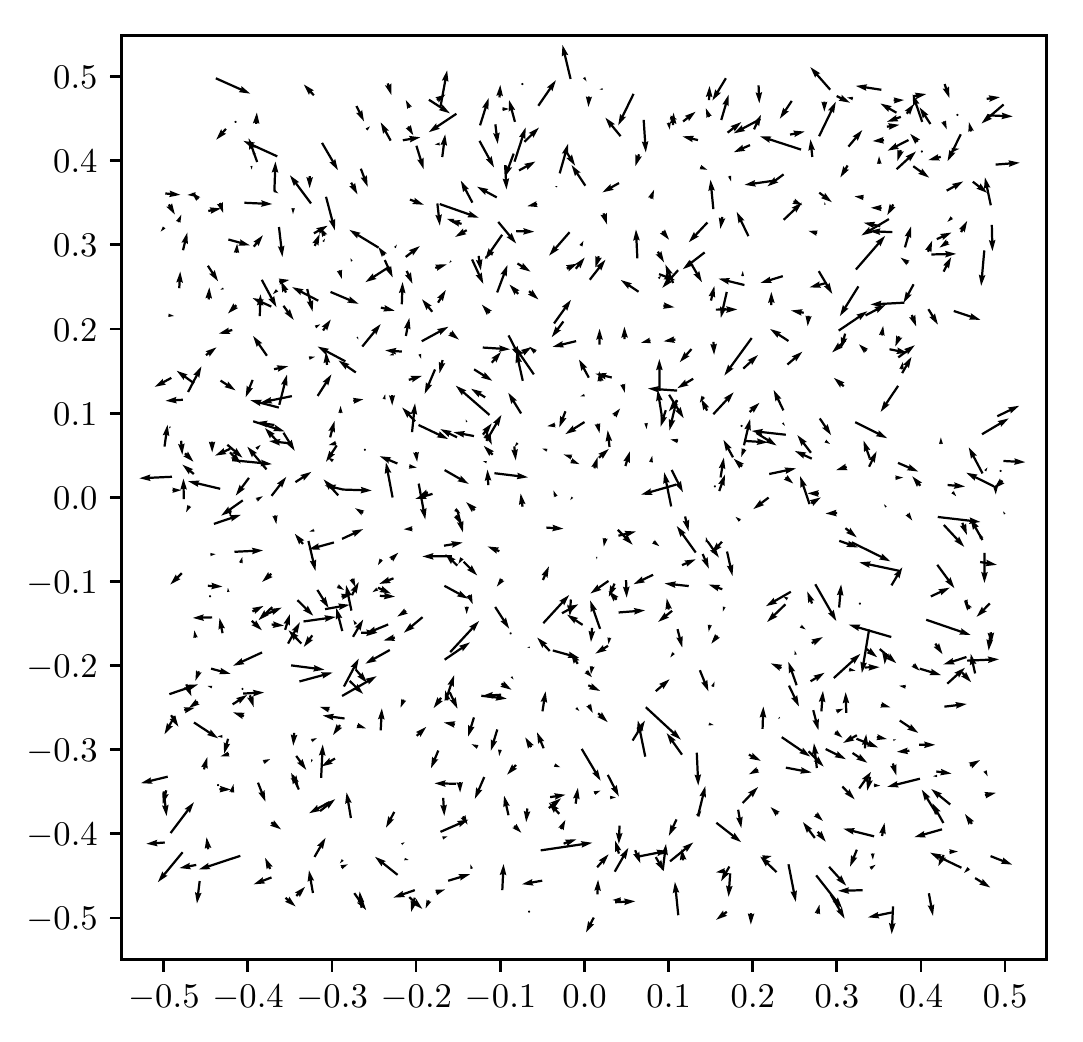}
  \caption{A simulated region with an $R_{\rm {sig}}$ of 0.07. Each star is shown as an arrow where
  the length and direction of the arrow reflects the star's velocity, and its position is the position of
  the star.}
  \label{expanding_region}
\end{figure*}

For reference a simulated region with $R_{\rm {sig}} = 0.07$ is shown in Fig. \ref{expanding_region}. By eye
the expanding nature of this region can not be readily observed. This again demonstrates
 the need for quantitative techniques to
analyse velocity substructure in star forming regions.

As will be discussed further in section \ref{comparison}
the velocity dispersion of Cygnus OB2 is difficult to characterise but
it is likely of order 10s of km s$^{-1}$. \citet{Wright16} finds a velocity dispersion of 17.8 km s$^{-1}$, assuming
this is correct VSAT should be able to distinguish expansion $\geqslant$ 1.2 km s$^{-1}$, therefore
any systematic expansion in the region must be below this threshold.

\section{Discussion}\label{discussion}

\subsection{Discussion of results using $\Delta v_{\rm M}$}\label{discuss_dv_mag}

The results of our VSAT analysis of this part of Cygnus OB2 can be summarised as follows:
\begin{itemize}
\item there is velocity structure on scales $<0.5$ parsecs;
\item scales $>0.5$ parsecs and $<5.5$ parsecs are consistent with a random velocity field;
\item reliable conclusions cannot be drawn at scales $>5.5$ parsecs.
\end{itemize}

In discussing these results it is also important to recall that the velocity dispersion
of the data is of order 10s of km s$^{-1}$, the size of the field of view is of order
10 parsecs, and the age of the region is of order a few Myrs \citep{Massey95, Wright15a}. As such is is likely the majority
of the stars in the dataset did not form in this field of view but moved here from their
birth sites of order 10s of parsecs away, erasing their original velocity structure. In this context the consistency of the results with a
random velocity field is to be expected as the distances between the stars relative to one another
represents a very small fraction of the total distances the stars have traveled from their origins.

Nevertheless there is velocity structure on scales $<0.5$ parsecs. This structure must either be
primordial or have developed over time. It is extremely difficult to see how such structures could
be constructed in a complex, unbound system such as a star forming region, so it is reasonable to assume
these structures are primordial, i.e. this signature is due to probably fairly low-$N$
groups\footnote{It is not clear to us if they deserve the title `clusters'}
of stars that formed near each other and so with similar velocities according to
the hierarchical paradigm of star formation \citep{Elmegreen00}.
We can also conclude these groups must be bound as in order for unbound stars
with similar velocities to remain within around 0.5 parsecs of each other
for at least 3 Myr (the approximate age of the region) the would have needed to form with a velocity
dispersion of $\lesssim$ 0.17 km s$^{-1}$.

To test this the local (within 0.5 pc) velocity dispersion of stars that are overrepresented in $<$ 0.5 pc bins are compared to
the velocity dispersion of those that are not and are found to be a factor of 2.09 smaller
on average. This provides further evidence of kinematic groups on 0.5 pc scales. However the velocity
dispersions of stars overrepresented in $<$ 0.5 pc bins are many times in excess of 0.17 km s$^{-1}$,
indicating the groups must be bound in order to have remained so closely spatially associated.
That said, it is likely that some stars have become unbound from the groups since they formed due
to dynamical interactions, and that these groups formed with somewhat higher $N$ and larger size. However
due to the youth of this region it is unlikely that the $N$ and size of these groups has changed
a great deal since their formation.

Observationally the hypothesis of primordial bound groups with sizes of order 0.5--1 parsecs
is supported by the fact this scale is the typical size of star clusters (see
e.g. \citet*{PortegiesZwart10}), the size of a typical molecular clump
\citep{Beuther07}, and is seen locally e.g. in Taurus \citep{Gomez93}.
A large-scale random velocity field with many `embedded' $\sim 0.5$ parsec
bound groups is also consistent with the high degree of spatial structure found
by \citet{Wright14} (they measure the \citet{Cartwright04} $Q$-parameter
to be 0.4--0.5).  It also fits the finding of \citet*{Griffiths18} that
the number of wide massive binaries in Cygnus OB2 suggests many (at least
30) different sites of massive star formation across the entire association.

\subsection{Discussion of results using $\Delta v_{\rm D}$}

As discussed in section \ref{results_dir} the velocity structure as measured using the directional
definition of velocity difference is very flat. $\Delta v_{\rm D}$ is consistent
with 0 for $\Delta r$ $<$ 3.5 parsecs, indicating the region is neither expanding or contracting on
these scales. Further, for $\Delta r$ between 3.5 and 6 parsecs fluctuations
in $\Delta v_{\rm D}$ are within the bounds of the cases where any
systematic velocity structure is removed by random shuffling
of the velocity vectors between stars (grey area in Fig. \ref{dv_dir}). As
such we conclude that any systematic expansion or contraction occurring within
the region at scales $<$ 6 parsecs is less significant than random noise. Further
VSAT is able to distinguish systematic expansion/contraction from random noise if the ratio
of the systematic component to the velocity dispersion of the region is $\gtrsim$ 0.07, so if any exists it must
be very small.

At larger $\Delta r$ there is a positive correlation between $\Delta v_{\rm D}$
and $\Delta r$. This indicates that at these scales stars further
away from each other tend to be moving apart, i.e.
the region may be expanding from the edges. However the VSAT method is not reliable
at such large $\Delta r$ because of small number statistics so an additional statistical test
is conducted and no expansion is found.

As a result we conclude from Fig. \ref{dv_dir} that there is no evidence of
systematic expansion or contraction on any scale
in Cygnus OB2, and if any exists it is far less significant than random noise.
This makes sense if, as we argue in section
\ref{discuss_dv_mag}, the stars in this dataset formed across the Cygnus star
forming complex and happen to be in this field of view at the time of observation. If this is the case we would
not expect the stars to move in systematic directions relative to one another, and indeed directional structure is not
observed in Fig. \ref{dv_dir}.

\subsection{Comparison of results to other kinematic studies of Cygnus OB2} \label{comparison}
As has been mentioned Cygnus OB2 has been the subject
of a number of other kinematic studies. Here we compare our
results to the findings of \citet{Wright14}, \citet{Wright16}, and \citet{Winter19}.

In section \ref{results_mag} we find evidence of kinematic substructure
at scales $\leqslant$ 0.5 pc. In agreement with this \citet{Wright16}
applies a number of statistics which confirm the presence
of kinematic substructure in the region. However from
a by-eye inspection of the data \citet{Wright16} asserts
that kinematic substructure is present at a range of scales
which is not consistent with our results.
From the results of section \ref{results_mag} we argue that
Cygnus OB2 formed with a hierarchical, substructured morphology.
This is in agreement with \citet{Winter19} which
uses $N$-body simulations and observations of the properties
of protoplanetary disks in Cygnus OB2 to constrain the initial conditions of
the region, and finds the region must have formed
with substructure to reproduce the observations.
Specifically \citet*{Winter19}
finds Cygnus OB2 may originate from a superposition of multiple clusters,
and their Figure 8(a) looks to the
eye to be moderately similar to our Fig. \ref{plot_of_region}.  However, it
might well lack the $0.5$ parsec structures we find in the real data (as
that scale was not imposed in their initial conditions and it is difficult
to see how it could arise dynamically).  A possible avenue for future work
could be to analyse a variety of different idealised initial conditions to
see which best-fit the real data.

We do a simple analysis and find the velocity dispersion
of the region to be 28.72 $\pm$ 0.62 km s$^{-1}$ which is not consistent
with the result found by \citet{Wright16} which
is 17.8 $\pm$ 0.6 km s$^{-1}$. This disagreement is explained by the
fact the datast used here covers a larger region than \citet{Wright16}.
This discrepancy notwithstanding in
both cases the region is found to be unbound. Additionally
\citet{Winter19} (which also looks at a larger region than \citet{Wright16}) estimated the initial velocity dispersion
of the region to be 50 km s$^{-1}$.

We conclude that the velocity dispersion is relatively
high and is likely of order 10s of km s$^{-1}$. Given this and the
region's age we conclude some `mixing' should have had time
to occur, erasing velocity substructure on scales that are not
bound. This fits uncomfortably with the assertion of \citet{Wright16}
that the region is not dynamically evolved. We
argue that given the relatively low-density nature of Cygnus
OB2 that we would not expect stars to have had a very large
number of dynamical interactions despite traveling potentially large distances.
Nevertheless we expect some degree of
dynamical evolution to have occurred.

In \citet{Wright16} the total kinetic energy of stars
moving radially inwards and outwards from the region are
compared and found to be nearly identical (the ratio is 51 per cent
to 49 per cent). From this it is concluded that there is no net expansion
or contraction in Cygnus OB2. This result is called into
question by \citet{Winter19} which argues that this ratio
does not provide a good measure of whether a region is expanding or contracting. However in
this paper, using an entirely different method, we also find that there is no evidence
of expansion or contraction in Cygnus OB2. This result suggests that the region
did not evolve to its current large, diffuse structure by expanding from a significantly denser one
as would be expected by some models \citep{Lada91}
\citep{Carpenter00}.

This suggestion is in agreement with the
conclusions drawn in \citet{Wright14} from their analysis
of Cygnus OB2's spatial structure.
Despite its consistency with previous results this finding
is still odd. The Cygnus OB2 region has a velocity dispersion
of 10s of km s$^{-1}$ and, given its estimated mass of $\sim$ 10$^{4}$ $M_\odot$,
it should be supervirial. Conventional logic states it should
be expanding rapidly, but it does not appear to be doing so.
However from its studies of protoplanetary disk properties
\citet{Winter19} estimates that gas removal ceased just
0.5 Myr ago and the region was not originally unbound. It
is conceivable that there has been insufficient time for the
region to `feel the effects' of this change. This could explain
why it is not noticeably expanding yet, but may begin to
do so in the near future. This would be consistent with the
conclusion drawn in \citet{Wright16} that within $\sim$ 4
Myr Cygnus OB2 will expand to be over 100 pc in size.

\section{Conclusions} \label{conclusions}

We analyse the positions and proper motions of stars in a region of Cygnus OB2 from \citet{Wright14} using the VSAT method presented in \citet{Arnold19}.

Our main findings are:
\begin{itemize}
  \item Stars within 0.5 parsecs of each other have significantly similar velocities.
  \item At all reliable scales larger than 0.5 parsecs velocities are consistent with a random distribution.
  \item We find no evidence of systematic expansion or collapse in this part of Cygnus OB2. If any such systematic motion exists its significance is $\lesssim$ a factor of 0.07 of that of the velocity dispersion of the region, and cannot be distinguished from random noise by the methods employed in this paper.
\end{itemize}

This suggests that we are observing many primordial bound structures on scales $<0.5$
parsecs (`groups' or small `clusters'). However, any initial velocity structure on
scales larger than 0.5 parsecs has been erased by stars having moved 10s of parsecs
in the few Myr since they formed.  Within this region we see no significant evidence
for global expansion or contraction (but we note this is only the central part of Cygnus OB2).

\section*{Acknowledgements}

NJW acknowledges an STFC Ernest Rutherford Fellowship (grant
number ST/M005569/1).
BA acknowledges PhD funding from the University of
Sheffield. Many thanks also to Stuart Littlefair for useful discussions and insights,
and the referee for helpful comments and suggestions which improved this work.

%%%%%%%%%%%%%%%%%%%%%%%%%%%%%%%%%%%%%%%%%%%%%%%%%%


\bibliographystyle{mnras}
\bibliography{mybibfile}

% Don't change these lines
\bsp	% typesetting comment
\label{lastpage}
\end{document}